\documentclass[twocolumn,nofootinbib,prd,showpacs,preprintnumbers,amsmath,amssymb]{revtex4}

\usepackage{amssymb}

\begin{document}

\title{Brane inflation and the fine-tuning problem}

\author{Abderrahim Chafik}
\affiliation{Laboratoire de Physique Th\'eorique, Facult\'e des 
Sciences,\\
BP 1014, Agdal, Rabat, Morocco,}

\date{\today}

\begin{abstract}

Brane inflation can provide a promissing framework for solving the fine-tuning problem in standard inflationary models. The aim of this paper is to illustrate the mechanism by which this can be achieved. By considering the supersymmetric two-stage inflation model we show that the initial fine-tuning of the coupling parameter can be considerably relaxed. SubPlanckian values of the inflaton during inflation can also be obtained. 

\end{abstract}

\pacs{ 98.80.Cq}

\maketitle

\section{Introduction}

Inflation has been initially invoked as a remedy for the conceptual problems of the cosmological hot big-bang model \cite{key1,key2}. These problems are essentially related to the fine-tuning of the initial conditions. However, inflation itself has given rise to another fine-tuning problem. Indeed, small values of the parameters of the theory are usually required in order to guarantee a sufficiently long period of inflation (i.e. flat potential) to solve the problems of the standard cosmological model (as the horizon and flatness ones), and to generate the correct magnitude of the density perturbations supposed to be the origin of the structure formation in the universe. This has been a great challenge for physicists.\\

Nevertheless, this problem can be removed in the context of the hybrid inflation \cite{key3,key4} which is one of the most attractive models of inflation, and is particularily relevent to the supersymmetric realisation of inflation \cite{key5} where a flat potential is naturally obtained. Hybrid inflation has also adressed a second issue which is the superPlanckian values of the inflaton field during inflation.\\

Unfortunately, these two problems reappear in a variant of hybrid inflation. Indeed, in reference \cite{key6} we have proposed a model of supersymmetric hybrid inflation which describes a scenario of two disconnected stages of inflation, but this can be only achieved with a severe fine-tuning of the coupling parameter. The same fine-tuning is required in order to generate the correct amplitude of  the density perturbations.\\

Recently there has been a great deal of interest in conceiving our universe to be confined in a brane embeded in a higher dimensional space-time \cite{key7}. Such models are motivated by superstring theory solutions where matter fields (related to open string modes) live on the brane, while gravity (closed string modes) can propagate in the bulk \cite{key8,key9,key10}. In these scenarios extra dimensions need not be small and may even be infinite \cite{key11,key12}. Another important consequence of these ideas is that the fundamental Planck scale $M_{4+d}$ in $(4+d)$ dimensions can be considerably smaller than the effective Planck scale $M_p = 1.2\times 10^{19}$ GeV in our four dimensional space-time.\\

It has also been noticed that in the context of extra dimensions and the brane world scenario the effective four-dimensional cosmology may deviate from the standard Big-Bang cosmology \cite{key13,key14,key15,key16,key17}, which could lead to a different physics of the early universe. The origin of this difference is the correction to the Friedmann equation which becomes \cite{key14,key15,key16,key17} 
\begin{equation}
H^2 = \frac{8\pi}{3M_p^2}\rho\biggl[1 + \frac{\rho}{2\lambda}\biggr] 
\end{equation}
where $\lambda$ is the brane tension. The new term in $\rho^2$ is dominant at high energies, but at low energies we must recover the standard cosmology, in particular during the nucleosynthesis, which leads to a lower bound on the value of the brane tension $\lambda\ge (1 MeV)^4$ . \\

With this modified Friedmann equation and within the slow-rolling paradigm the two slow-rolling parameters become \cite{key18}
\begin{eqnarray}
\epsilon & \equiv & \frac{M_p^2}{16\pi}\biggl(\frac{V'}{V}\biggr)^2\biggl[\frac{2\lambda(2\lambda + 2V)}{(2\lambda + V)^2}\biggr] \\
\eta & \equiv &  \frac{M_p^2}{8\pi}\biggl(\frac{V''}{V}\biggr)\biggl[\frac{2\lambda}{2\lambda + V}\biggr]  
\end{eqnarray}

where, as usually assumed in the slow-roll approximation, the energy density is dominated by the potential energy $V$. It is clear that at low energies, $V\ll\lambda$, the two parameters reduce to the standard form \cite{key1}, but at high energies, $V\gg\lambda$, this extra contribution to the Hubble expansion helps damp the rolling of the scalar field as $\epsilon$ and $\eta$ are suppressed. Thus, {\it brane effects ease the condition for slow-roll inflation for a given potential} \cite{key18}.\\

In the same way the number of e-folds during inflation become \cite{key18} 

\begin{equation}
N \simeq -\frac{8\pi}{M_p^2}\int_{\sigma_1}^{\sigma_2} \frac{V}{V'}\biggl[1 + \frac{V}{2\lambda}\biggr]d\sigma 
\end{equation}

Similiraly, at high energies this expression yields more inflation between any two values of the inflaton for a given potential. It follows from these two remarks that we can hope to solve or to reduce the fine-tuning problem in the inflationary models in the context of the braneworld scenario since as it has been mentioned, this problem is essentially due to the requirement of a sufficiently long period of inflation.

\section{The initial model}

In reference \cite{key6} we have contributed to the resolution of the problem encountered in the first realisation of the hybrid inflation model in supersymmetry: the generation of a slope for the scalar potential. Indeed, the initial version of the supersymmetric hybrid inflation was based on a superpotential of the form \cite{key19,key20}
\begin{equation}
W = \kappa S(-\mu^2 + \overline{\phi}\phi)
\end{equation}
where $\overline{\phi}$, $\phi$ is a conjugate pair of superfields transforming as non-trivial representation of some gauge group $G$ under which the superfield $S$ is neutrally charged, the coupling parameter $\kappa$ and the mass scale $\mu$ can be taken to be real and positive.\\

Such a superpotential gives a scalar potential which is identical to the potential of the initial non supersymmetric hybrid model \cite{key3,key4} but without the mass-term of the inflaton field . Such a mass-term is necessary for inflation since it gives the slope of the valley of minima, and then drives the inflaton to its critical value where inflation ends.\\

Our approach was based on the modified superpotential \cite{key6}
\begin{equation}
W = \kappa S(-\mu^2 + \overline{\phi}\phi)\; +\; \frac{\lambda_S}{3}S^3
\end{equation}

From the above superpotential we have derived an effective potential of the form
\begin{equation}
V(\varphi\; ,\;\sigma)\;\;=\;\;\kappa^2\biggl(\mu^2\; -\;\frac{\varphi^2}{4}\biggr)^2\; +\; g^2\frac{\varphi^2\sigma^2}{4}\; +\;\lambda_{\sigma}\frac{\sigma^4}{4}
\end{equation}
(where $\lambda_{\sigma}$=$\lambda^2_S$) which is the same as the one of the initial Linde's model but the mass-term has been replaced by a quartic self-coupling one\footnote{The scalar fields $\varphi$ and $\sigma$ are proportional to the real parts of the scalar components of the superfields $\overline{\phi}$, $\phi$ and $S$ respectively .} .\\

In the resulting model a first phase of inflation occurs where the potential is dominated by the $\sigma^4$-term, and ends when the inflaton reaches a value $\sigma_e\sim {\cal O}(M_p)$ ($\epsilon(\sigma_e)\sim 1$) which is much greater than $\sigma_c\; =\;\sqrt{2}\kappa\mu/g$ . The inflaton undergoes a phase of oscillations about its minimum during which its energy decreases untill the vacuum energy $V\; =\; \kappa^2\mu^4$ becomes dominant, then a second stage of inflation begins. The achievement of such a scenario requires the condition :\\

\begin{equation}
\lambda_{\sigma}\frac{\sigma^4_{60}}{4}\;\ll\;\kappa^2\mu^4
\end{equation}
where $\sigma_{60}$ is the value of $\sigma$ 60 e-folds before the end of inflation, which is given by the expression \cite{key6}
\begin{equation}
\frac{1}{\sigma^2_{60}}\;\; =\;\;\biggl[\frac{1}{\kappa^2\mu^2}\biggl(\frac{g^2}{2}\; -\; \frac{30}{\pi}M^2_p\frac{\lambda_{\sigma}}{\mu^2}\biggr)\biggr]
\end{equation}

Substitute Eq.(9) into Eq.(8) yelds
\begin{equation}
\lambda_{\sigma}\;\;\simeq\;\; 5\times 10^{-12}
\end{equation}

The same value of the coupling parameter has also been deduced from the observational constraint $\delta_H\; =\; 1.9\times 10^{-5}$.\\

This model describes a cosmologically rich scenario since the first inflationary phase may produce the homogeneity beyond the Hubble radius which makes natural the onset of the second stage during which the interesting length scales leave the horizon. However, a severe fine-tuning of the self-coupling constant is required to achieve such a second stage and to generate the correct magnitude of the spectrum of density perturbations.
\section{The brane version}
The above result may be interpreted as reflecting the simple fact that the two-stage scenario is less favourable and thus we should only consider the case of a single epoch of inflation where the $\sigma^4$ term remains dominant untill the instability point $\sigma=\sigma_c$, which is identical to the original model of chaotic inflation \cite{key21}. However, even in this case a fine-tuning is inevitable as was shown in in reference \cite{key22} where the COBE constraint has given the standard result :
\begin{equation}
\lambda_{\sigma}= 4.2\times 10^{-13}
\end{equation}

As it was explained in the introduction this situation can be improved in the context of the braneworld scenario, where the fine-tuning will be considerably relaxed in the second case.\\

In this section we shall restrict ourselves to the simple picture of the so-called Randall-Sundrum II model \cite{key23} where the matter of the universe is confined in a (3+1) brane with positive tension (the bulk space beeing empty), and the higher dimensional space is non compact. The fundamental equations in this case are Eq.(1)-(4). Hence, by calculating the value of $\sigma_{60}$ using the expression Eq.(4) of the number N we find
\begin{equation}
\frac{1}{\sigma^2_{60}}\;\; =\;\;\frac{g^2}{2\kappa^2\mu^2}\; -\; \frac{15M^2_p}{\pi}\;\frac{\lambda_{\sigma}}{\kappa^2\mu^4}\biggl[1\; +\;\frac{\kappa^2\mu^4}{\lambda}\biggr]^{-1}
\end{equation}

The condition Eq.(8) then becomes
\begin{equation}
\frac{g^2}{\kappa\lambda_S}\;\; \gg\;\; \frac{30M^2_p}{\pi}\;\frac{\lambda_S}{\kappa\mu^2}\biggl[1\; +\;\frac{\kappa^2\mu^4}{\lambda}\biggr]^{-1}
\end{equation}
or equivalently
\begin{equation}
g^2\mu^2\biggl[1\; +\;\frac{\kappa^2\mu^4}{\lambda}\biggr]\;\; \gg\;\;\lambda_{\sigma} M^2_p
\end{equation}

In the brane inflation context the main cosmological constraint comes from the amplitude of the scalar perturbations which is given by \cite{key18}:
\begin{equation}
A_s^2 \simeq \biggl(\frac{512\pi}{75M_p^6}\biggr)\frac{V^3}{V'^2}\biggl[\frac{2\lambda + V}{2\lambda}\biggr]^3\Biggl\vert_{k = aH} 
\end{equation}

In our model we obtain
\begin{equation}
A_s^2\;\;\simeq\;\; \biggl(\frac{512\pi}{75M_p^6}\biggr)\biggl(\frac{g^3\mu^3}{\lambda_{\sigma}}\biggr)^2\biggl[1\; +\;\frac{\kappa^2\mu^4}{\lambda}\biggr]^3
\end{equation}
where we have used the condition Eq.(13) to neglect small terms.\\

The COBE observed value of $A_s$ is
\begin{equation}
A_s\; =\; 2\times 10^{-5}
\end{equation}

Using Eq.(13) the above value implies
\begin{equation}
\lambda_{\sigma}\;\;\ll\;\; 2\times 10^{-12}
\end{equation}
which implies that even in the braneworld scenariothe the two-stage inflation requires fine-tuning. However, in the case of the simple hybrid inflation with a quartic term the result is very interesting. Indeed, if the $\sigma^4$ term remains dominant the value of $\sigma_{60}$ is given by:

\begin{equation}
\sigma_{60}^2\;\;\simeq\;\;\frac{1440}{\pi}M^2_p\frac{\lambda}{\lambda_{\sigma}}\; +\; 8\biggl(\frac{\kappa\mu}{g}\biggr)^6
\end{equation}
and then the amplitude of the scalar perturbations becomes
\begin{equation}
A^2_s\;\;\simeq\;\;\frac{8\pi}{75}\lambda_{\sigma}\biggl(\frac{180}{\pi}\biggr)^3
\end{equation}
where we have used the upper bound of the brane tension \cite{key24}
\begin{equation}
\lambda\;\;\leq\;\; (1.7\times 10^{-4})^4\; M^4_p
\end{equation}

The new result is then
\begin{equation}
\lambda_{\sigma}\;\;\simeq\;\; 0.8\times 10^{-6}
\end{equation}
or equivalently $\lambda_S\simeq 0.9\times 10^{-3}$ for the fundamental parameter of the theory Eq.(6).\\

With this value of the coupling parameter $\lambda_{\sigma}$ the fine-tuning in the initial model has been clearly relaxed as was expected. Furthermmore, as we have also expected, the value of the inflaton field during inflation can be made less than the Planck mass. Indeed, using Eq.(4) the condition $N\geq 70$ for a sufficient inflation (with $\sigma_e\equiv \sigma_c$) gives the following initial value
\begin{equation}
\sigma_i\;\; \geq 4\times 10^{-2}\; M_p
\end{equation}

Nevertheless, the two-stage scenario is not completely lost but its achievement with the present value of the coupling paramter [Eq.(22)] requires superPlanckian values of the inflaton. Indeed, as a minimal value of $\sigma$, in this case, the value $\sigma_e >\sigma_c$ which satisfy $\epsilon(\sigma_e)\sim {\cal O}(1)$ is found to be $\sigma_e\simeq 16\; M_p$.

\section{Discussion and conclusion}
In this paper the model of supersymmetric two-stage inflation has been reexamined in the context of the braneworld scenario. The aim of this extention was to relax the fine-tuning of the coupling parameter in the initial version. This goal has benn achieved but without having two disconnected stages of inflation, or by keeping superPlanckian initial values of the inflaton field. The resulting scenario is then a simple hybrid inflation with a quartic term where the potential is dominated by a $\sigma^4$-term. Such a model also requires fine-tuning in the standard case. In addition to this result it has been shown that the initial value of the inflaton field can be less than the Planck mass as it was initially achieved in the Linde's hybrid inflation. So we have illustrated how the fine-tuning, usually seen as a generic problem of the standard inflationary models, can be relaxed in the braneworld picture.\\

The possibility of solving the fine-tuning problem in extra dimensions theories has also been studied in reference \cite{key25} where the authors have proposed a generalization of the so-called assisted inflation \cite{key26} (see also \cite{key27}) for power law potentials where the Kaluza-Klein modes of the 5-dimensional scalar field constitute the source of the necessary multiplicity of scalar fields. In this model the 4-dimensional coupling constant of the K-K fields is determined by the 5-dimensional one divided by the number of the K-K modes. As a result the effective coupling constant is suppressed by the number of scalar fields present in the theory, which means that it becomes naturally small (to satisfy the COBE constraint) without the need of any fine-tuning of the fundamental coupling constants. In the same framework the authors have also adressed the issue of the large initial values of the inflaton field in chaotic inflation \cite{key28}. They have demonstrated that for the initial values of the constituent fields many order of magnitude smaller than $M_p$ the value of the inflaton can exceed the threshold of few $M_p$. While through this mechanism the solution of the above two problems depends on the number of the scalar fields present in the theory, in the braneworld scenarion it is directely related to the extra dimensions aspect of the cosmological equations. \\

On the other hand, the hierarchy problem which is usually seen as a fine-tuning problem can also be addressed in the brane theory \cite{key29}, where in the so-called Randall-Sundrum model I with two branes the very small rapport between the electroweek and the Planck scales is due to the highly curved AdS background, which implies a large gravitational red-shift between energy scales in the two branes. In this model it is shown that by an appropriate inter-brane distance the 5-dimensional Planck scale $M_5$ is not very far from the electroweek scale. In the same framework there has been also a new hope to solve the cosmological constant problem, in particular through the so-called self-tuning mechanism (see e.g. \cite{key30} and references therein). We then conclude that the extra dimension theories, and in particular the brane theory, are very promessing for the solution of the challenging fine-tuning problem.

\begin{acknowledgments}
I would like to thank the Abdus Salam ictp for hospitality.
\end{acknowledgments}

\bibliography{art4.bib}

\end{document}